\documentstyle[12pt,epsfig]{article}
\newcommand{\blankline}{\vskip .3cm}
\newcommand{\f}{\begin{equation}}
\newcommand{\ff}{\end{equation}}
\begin{document}
\centerline{\LARGE  $\cal M$ theory as a matrix extension of 
Chern-Simons theory}
\blankline
\centerline{Lee Smolin${}^*$}
\blankline
\centerline{\it  Center for Gravitational Physics and Geometry}
\centerline{\it Department of Physics}
\centerline {\it The Pennsylvania State University}
\centerline{\it University Park, PA, USA 16802}
\centerline{and}
\centerline{\it The Blackett Laboratory}
\centerline{\it Imperial College of Science, Technology and Medicine}
\centerline{\it South Kensington, London SW7 2BZ, UK}
\blankline
\blankline
\centerline{Jan 23, 2000}
\blankline
\blankline\blankline
\blankline
\centerline{ABSTRACT}
We study a new class of matrix models, the simplest of which
is based on an $Sp(2)$ symmetry and has a compactification which is equivalent
to Chern-Simons theory on the three-torus.  By replacing $Sp(2)$ 
with the  super-algebra $Osp(1|32)$, which has been conjectured to be the 
full symmetry group  of $\cal M$ theory, we arrive at a supercovariant matrix 
model which appears to contain within it the previously 
proposed $\cal M$ theory matrix models.  There is no background spacetime so that 
time and dynamics are introduced via compactifications
which break the full covariance of the model.
Three compactifications are studied corresponding to a hamiltonian
quantization in $D=10+1$, a Lorentz invariant quantization in 
$D=9+1$ and a light cone gauge quantization in $D=11=9+1+1$.
In all cases constraints arise which eliminate certain  
higher spin fields in terms of lower spin dynamical fields.  In the
$SO(9,1)$ invariant compactification we argue that the one loop 
effective action reduces to the IKKT covariant matrix model.  In the light cone
gauge compactification the theory contains the standard $\cal M$
theory light cone gauge matrix model, but there appears an 
additional transverse five form field.
\vfill
${}^*$smolin@phys.psu.edu
\eject

\section{Introduction}

In this note we present a possible approach to $\cal M$ 
theory based on a simple matrix model, which describes the
dynamics of a matrix which is built from the super Lie
algebra $Osp(1|32)$.   The original motivation for studying this
model came from an attempt to simplify a proposal for
a background independent formulation of $\cal M$ theory\cite{mpaper}, 
based on a background independent approach to
a causal membrane field theory\cite{tubes,pqtubes}.
However the model which emerged is quite simple, and so
merits a separate presentation. The goal of the present paper 
is only to initiate a study of this model,  much more remains to be
done to understand its possible relationship to $\cal M$ theory.

The main motivation for this model is that it fully realizes the
supersymmetry algebra $Osp(1|32)$ which has been proposed by different 
authors as the ultimate symmetry group of $\cal M$ theory\cite{m}.  
As we shall argue, various matrix models and string theories may arise
from different  compactifications of the model.
A second motivation was to find an extension of the dWHN-BFSS
matrix model\cite{CH,dWHN,BFSS,moreBFSS},which incorporates the full $10+1$
dimensional
super-Poincare invariance of the super-membrane and the flat
space limit of $11$ dimensional supergravity.   A third motivation was
to find a single model which includes both that light cone gauge
theory and the $SO(9,1)$ supercovariant IKKT matrix model\cite{IKKT} as
different reductions.

It is not obvious that such an extension of the matrix model should
exist.  Among the reasons to think it should not are that such a 
theory is not likely to be related to either $10$ dimensional 
super-Yang-Mills theory or the $11$ dimensional membrane, as it can be
shown \cite{memembrane} that once light cone gauge has been lifted the
gauge group of the membrane is too large to represent manifestly in
terms of the $N \rightarrow \infty$ limit of an $SU(N)$ gauge invariance.
A  perhaps even more serious issue is that covariantizations
of supersymmetric theories which realize the full supersymmetry 
linearly, and off shell, are normally plagued with ghosts and higher 
spin fields

Mindful of these potential pitfalls, we proceed here to invent and study a model.
The model differs from perviously studied matrix models in that the 
action is derived by an extension of a matrix form of Chern-Simons 
theory which we describe in the next section.  This action is cubic 
in the matrices.  One might worry 
that this leads to instabilities, however as in the case of pure 
general
relativity (with vanishing cosmological constant and in the compact 
case) the action vanishes
on shell.  In fact the theory 
shares several characteristics with first order 
formulations of general relativity and supergravity, some of which
are also  cubic in the basic variables\cite{CDJ,hologr,superholo}.  
In those theories time is only introduced by expanding the theory around
a particular classical background.  Since the action is cubic
this leads to an expression first order in time derivatives.  Thus, 
each choice of time leads to a phase space description.  When the
canonical theory is analyzed it is found that
there are always constraints which resolve the possible problems of
ghosts and higher spin fields, leading in the end to a sensible theory,
at least classically. Below we will show that the cubic matrix
action defines a theory with similar characteristics. In particular 
time is only introduced when the theory is expanded around particular
classical solutions that define a compactification.  We also find that
constraints arise which eliminate higher spin fields in terms of 
lower spin fields.

In the next section we study a  simple theory with a cubic action 
in which the matrices are valued in 
$Sp(2)$.  We show how compactifying it on a circle defines a phase 
space and that when it is compactified on the three-torus
it is equivalent to Chern-Simons theory. 
In section 3 we extend that model 
simply by replacing $Sp(2)$ by the superalgebra $Osp(1|32)$.  The rest
of the paper is then devoted to the study of this model.
In section 4 we
perform a hamiltonian analysis relevant for a $10+1$ dimensional
quantization of the theory and we find that there are constraints
which eliminate many of the degrees of freedom. In sections 5 and 6
we study respectively compactifications that reduce the symmetry to
the super-Poincare group in $9+1$ dimensions and the super-Euclidean
group in $9$ dimensions. Because of the constraints we are unable to
make a precise computation of the effective action, however we
are able to argue from symmetry that in the first case the covariant
matrix model proposed by IKKT \cite{IKKT} is reproduced.  In the
second case, by going to light cone gauge in $D=11$ we arrive at
a theory that contains the standard dWHN-BFSS matrix model relevant
for the light cone gauge description of $\cal M$ theory in flat
$10+1$ dimensional spacetime\cite{CH,dWHN,BFSS}.  
It seems in addition to contain
one more field, which is a transverse five form field.  

\section{Matrix representation of topological field theory}

We will take as our starting point a fundamental fact about general
relativity and supergravity, which is that they arise by constraining
the actions for topological quantum field 
theories\cite{CDJ,hologr,superholo}.  This suggests
the following strategy: find a way to represent some topological
quantum field theory as a matrix model, and then find a way to
naturally extend it to include the symmetries $\cal M$ theory is
expected to have.

What form of an action shall we  use? It is clear that if
we use a conventional matrix theory action involving quadratic
and quartic terms we will not get a representation of a 
topological field theory. Furthermore, when we extend the symmetry
to a covariant superalgegbra there will be a great danger of ghost fields and
negative norm states coming from the minus signs in the spacetime
metric.  To avoid this we  study instead an action cubic
in the matrices.  Under compactification to define a time coordinate
this can produce an action at most first order in time derivatives.
Thus, such as action will define and live on a phase space.
This is attractive as first order, phase
space actions are a very convenient starting point for 
analyzing theories with spacetime gauge invariances.
They are also the starting point for the discovery of connections
between topological quantum field theory and gravitational theories.

To see the effect of using a cubic action, we may
consider a very simple model based on $Sp(2)$. Here the
field is given by a matrix
\f
M_{I}^{\ J}= \left [ 
\begin{array}{cc}
   x_{3}  & x_{1}  \\
    x_{2} & -x_{3}
\end{array}
\right ]
\ff
where $x_{i}$, $i=1,2,3$ are three  $N\times N$ matrices.
A simple cubic action is then given by
\f
I^{Sp(2)}= Tr \{ M_{I}^{\ J} [M_{J}^{\ K}, M_{K}^{\ I}] \} 
= 6 Tr \{ x_{1}[x_{2},x_{3}] \} 
\ff
where the trace and commutator are in the $N\times N$ matrix variables.
We see that in the classical theory they
must all commute with each other.  To introduce a time variable
we break the $Sp(2)$
invariance by  expanding around a vacuum given by $x_{3}=\hat{D}=\hat{d}+a_{3}$,
where, using the standard matrix compactification 
trick\cite{trick,BFSS} 
(to be recalled below), the action reduces in the limit $N\rightarrow 
\infty$ to
\f
I^{Sp(2)}=6 \oint dt Tr \{  x_{2}(t) \dot{x}_{1}(t)  + 
a_{3}(t)[x_{1}(t),x_{2}(t)]\}
\ff
where $x_{1}(t)$ and $x_{2}(t)$ are now two one parameter families of
$M\times M$ matrices and $a_{3}(t)$ is a one dimensional $gl(M)$  
gauge field. If we require that the matrices be hermitian,
we reduce this to a $U(M)$ gauge invariance.  
We see that the theory describes a phase space 
$\Gamma= (p,x_{1})$ with $p=\delta I / \delta \dot{x}_{1}=x_{2}$
with the constraint that as $M\times M$ matrices,
$[p(t), x_{1}(t)]=0$.  
In this simple case there is no dynamics, the theory is something
like a matrix version of a topological field theory.  

In fact the cubic form of the action is closely related to
topological field theory.  To see this let us consider
the same $Sp(2)$ model, but let us make a triple
compactification defined by the expansion
\f
x_{i}=\hat{D}_{i}=\hat{\partial}_{i} +a_{i}
\ff
where $\hat{\partial}_{i}$ are 3 $M\times M$ matrices that each give
a compactification and $[\hat{\partial}_{i},\hat{\partial}_{j}]=0$
so that when $a_{i}=0$ we have a solution to 
the classical equations of motion.  The cubic action is then equal in 
the limit to
\f
I^{Sp(2)}=3 \int_{T^{3}} d^{3}x \epsilon^{ijk} \ 
Tr \{ a_{i} \partial_{j}a_{k}+{1\over 3} a_{i} [a_{j}, a_{k}]\}
\ff
This is the action for $U(M)$ (or, with unconstrained matrices,
$gl(M)$) Chern-Simons theory on the three-torus.
The equations of motion are  $F_{ij}\equiv 
[\hat{D}_{i},\hat{D}_{j}]=0$.  
Thus the symplectic matrix model suggests that there is a connection between
a $2M$ dimensional phase space  and a $U(M)$ Chern-Simons theory.  The 
Chern-Simons theory may  be regarded as a gauge theory of 
the sympectic structure, so that the original phase space structure
is coded in the Poisson brackets amongst Wilson loops,
$W_{i}^{n}=Tr\{ P[e^{\oint dx_{i}a_{i}}]^{n} \}$, for $i=1,2$ and 
$n=1,..,M$.  We will not pursue
this further here, but go on to see how an extension of this cubic
sympectic matrix model may have something to do with $\cal M$ theory.

\section{The model}

We now extend the cubic matrix model by extending the $Sp(2)$
symmetry to the superalgebra
$Osp(1|32)$ which is believed to be the symmetry group of $\cal M$ 
theory. 
The degree of freedom of our theory will then be 
a set of unconstrained $N\times N$ matrices, each element of 
which is also valued in the adjoint representation of the superalgebra
$Osp(1|32)$.  
We first define the notation that we will use to describe the 
matrices that define the adjoint representation of
$Osp(1|32)$.  
These are $33 \times 33$ component matrices, $W_{\alpha}^{\ \beta}$
which are defined to satisfy  
\f
W\cdot G = - G \cdot W^{T} , 
\label{ospdef}
\ff
where $T$ stands for the matrix transpose and $G$ is the
$33 \times 33$ matrix,
\f
G_{\alpha}^{\ \beta} = 
\left [
\begin{array}{ccc}
    0 & -I & 0  \\
    I & 0 & 0  \\
    0 & 0 & 1
\end{array}
\right ]
\ff
Here the first two rows and columns are $16 \times 16$ dimensional
and the third row and column has one component.  We may then
write for the $33$ fold indices $\alpha, \beta,\ldots$,
$\alpha = A, A^{\prime}, 0$ where $A=1,\ldots, 16$
and  $A^{\prime}=1^{\prime},\ldots, 16^{\prime}$. The
$A$ and $A^{\prime}$ components have an even grading, so
$g(A)=g(A^{\prime}) =0$ while the $33$'d, $0$ component has
an odd grading, $g(0)=1$.

It is easy to see that the solutions to (\ref{ospdef})
may be parameterized as
\f
W_{\alpha}^{\ \beta} = 
\left (
\begin{array}{ccc}
    A & B & \Psi  \\
    C & -A^{T} & \Phi  \\
    \Phi^{T} & -\Psi^{T} & 0
\end{array}
\right )
\label{param1}
\ff
where $A$ is a $16 \times 16$ matrix,  $B$ and $C$ are 
$16 \times 16$ symmetric matrices, and $\Psi^{A}$ and
$\Phi^{A^{\prime}}$ are $16$ component spinors.
All quantities are real-Grassman valued, $A,B$ and $C$
are real even Grassman variables while  $\Psi^{A}$ and
$\Phi^{A^{\prime}}$ are real odd Grassmann variables.
It will be useful also to decompose $A^{AB}$ into its symmetric
and antisymmetric parts,
\f
A^{AB}= X^{AB}+ Y^{AB},
\label{param2}
\ff
where $X^{AB}= X^{(AB)}$ and $Y^{AB}= Y^{[AB]}$.

Let us now promote each component of $W_{\alpha}^{\ \beta}$
to an $N\times N$ matrix, which we will call
$Z_{\alpha a}^{\ \beta b}$, with $a,b,c,\ldots = 1,\ldots, N$.

We then define our theory by the cubic action,
\f
I= {1 \over g^{2} } Tr \left \{ Z_{\alpha }^{\ \beta } \left [ 
Z_{\beta }^{\ \gamma } , Z_{\gamma }^{\ \alpha } \right ]
\right \}
\label{cubicaction}
\ff
The (super)trace and the (super)commutator are both taken in the
$N$ component indices.  The supercommutator is defined by the
usual formula, for two grassman valued $N\times N$ matrices, $X$ and $Y$,
$[X,Y] \equiv XY - (-1)^{g(X)g(Y)}YX$.  We will call this the cubic
action in the rest of the paper.

The action (\ref{cubicaction}) has the following symmetries:
a) Global (that is commuting with $GL(N,R)$) supersymmetry
in which the components of $Z_{\alpha }^{\ \beta }$ transform
under the adjoint representation of $Osp(1|32)$.  b) Global
$GL(N,R)$ symmetry, c) a generalized translation symmetry, under
which
\f
Z_{\alpha a}^{\ \beta b}= Z_{\alpha a}^{\prime  \beta b} =
Z_{\alpha a}^{\ \beta b} + \delta_{a}^{b}V_{\alpha }^{\ \beta }
\ff
Note that the fields and the coupling constant $g$ are all
dimensionless, which of course is required as there is nothing
in the theory that refers to space or time.  We will generally
set $g=1$ for convenience.

One might make the following objections to this model.  First the action 
is not bounded from below or above, second there is no explicit
time coordinate, third there is a global translation symmetry.  The
first two are properties of general relativity so we should perhaps
not be surprised to see them in any theory that has general relativity
as a limit. Futhermore, we can point out that as in general relativity
in the compact case the action vanishes on solutions.
To introduce time we will have to expand the 
theory around a suitably chosen classical background.  (This by the 
way, agrees with some\cite{julian-time}, but not 
all\cite{causal,lotc}, views on the role of 
time in quantum gravity.)  Once time is defined in this way a 
Hamiltonian may be constructed. What is required is only 
that some of the theories 
defined by these hamiltonians are stable, for physically interesting
choices of backgrounds.  

The existence of a global translation symmetry is, however, not a 
feature of classical general relativity; it suggests that some 
background dependence has been left in, which should arise only
in the presence of certain classical solutions. It does however agree 
with some proposals concerning $\cal M$ theory in which the translations 
symmetry of flat $11$ dimensional spacetime is to be absorbed into a
larger symmetry group which sometimes has been proposed to be
$Osp(1|32)$\cite{m}.  

To answer this last criticism one can reduce the translation
symmetry by dropping the commutator, so that we have
\f
I^{gauged}= {1 \over g^{2} } Tr \left \{ Z_{\alpha }^{\ \beta }  
Z_{\beta }^{\ \gamma } Z_{\gamma }^{\ \alpha }  
\right \}
\label{gcubicaction}
\ff
We will refer to this as the gauged cubic action.  It has less
global symmetry, but a far larger gauge symmetry group, which is
given by the possibility of making $a$ valued $Osp(1|32)$ 
transformations.  This makes the model harder to analyze
although perhaps more interesting.  It will be discussed elsewhere.

Returning to the cubic action, the classical equations of
motion that follow from (\ref{cubicaction}) are simply
\f
\left [ Z_{\beta }^{\ \gamma } , Z_{\gamma }^{\ \alpha } \right ] =0
\label{eom}
\ff
To proceed to analyze the theory we need to decompose the
action in terms of (\ref{param1},\ref{param2}), this gives us
\begin{eqnarray}
I& = &{1 \over g^{2} } Tr \left \{
6X_{AB}[B^{BC},C_{C}^{A}] +  
6 Y_{AB}[X^{BC},X_{C}^{A}]+ 2 Y_{AB}[Y^{BC},Y_{C}^{A}] \right.
\nonumber  \\
&& \left. + 2 X_{AB}\{ \Psi^{A},\Phi^{B}\} + B_{AB}\{ \Phi^{A},\Phi^{B}\}
- C_{AB}\{ \Psi^{A},\Psi^{B}\}
\right \}
\end{eqnarray}

We will now discuss how the theory behaves when expanded around
three different backgrounds, which are solutions to the classical
equations (\ref{eom}).

\section{The first compactification and a Hamiltonian formulation}

In order to study the dynamics of the model we have to introduce a
time coordinate.  This can be done by the usual trick of choosing a
background which corresponds to an $S^{1}$, which  is interpreted as a
compactification of the model.  This necessarily breaks the symmetry of
the model to a subalgebra of $Osp(1|32)$.  But of course in a 
relativistic theory symmetry reduction is always a consequence of 
a choice of the time coordinate.

To see how to choose a time direction in the 
$Osp(1|32)$ version of the theory we may choose a 
coordinatization which describes an embedding of the $11$ 
dimensional Super-Poincare algebra in $Osp(1|32)$.  We do this
by choosing a real $32$ dimensional representation of $Cliff(10,1)$,
given by the following choice:
\f
\Gamma^{i}= \left (
\begin{array}{ccc}
    \gamma^{i} & 0 & 0  \\
    0 & -\gamma^{i} & 0  \\
   0 & 0 & 0
\end{array}
\right ) \ ; \ \Gamma^{10}=\Gamma^{\#}=\left (
\begin{array}{ccc}
    0 & I & 0  \\
    I & 0 & 0  \\
    0 & 0 & 0
\end{array}
\right ) \ ; \ \Gamma^{0}=\left (
\begin{array}{ccc}
    0 & -I & 0  \\
    I & 0 & 0  \\
    0 & 0 & 0
\end{array}
\right ) 
\ff
where $\gamma^{i}$ are $16 \times 16$, symmetric, real, nine dimensional
$\gamma$-matrices normalized by
$\gamma^{i}\gamma^{j}+\gamma^{j}\gamma^{i}= +2 \delta^{ij}$,
with $i=1,\ldots,9$.   It will be useful to note also the
corresponding representation of $Spin(10,1)\in Cliff_{0}(10,1)$,
\f
\Gamma^{ij}= \Gamma^{i}\Gamma^{j} = \left (
\begin{array}{ccc}
    \gamma^{ij} & 0 & 0  \\
    0 & \gamma^{ij} & 0  \\
   0 & 0 & 0
\end{array}
\right ) \ ; \ \Gamma^{\# i}= \left (
\begin{array}{ccc}
    0 & -\gamma^{i} & 0  \\
    \gamma^{i} & 0 & 0  \\
    0 & 0 & 0
\end{array} \right )
\ff
\f
\Gamma^{0 \#}=  \left (
\begin{array}{ccc}
    I & 0¥ & 0  \\
    0 & -I & 0  \\
    0 & 0 & 0
\end{array}
\right ) \ ; \ \Gamma^{0i}= \left (
\begin{array}{ccc}
    0 & \gamma^{i}¥ & 0  \\
    \gamma^{i}¥ & 0 & 0  \\
    0 & 0 & 0
\end{array}
\right ) 
\ff
Note that $\gamma^{ij}_{AB}$ is real and antisymmetric so that
$\Gamma^{ij}\in Sp(32)$.

In order to understand the physical content of the theory 
it is useful to understand the decomposition of the adjoint rep
of $Sp(32)$ into irrep's of $Spin(9)$.  This helps because 
$Spin(9)$ governs the degrees of freedom of the light cone gauge
of the $11$ dimensional theory, which is where the degrees of freedom
should be manifest and we expect to make contact with the 
standard $\cal M$ theory matrix model.
We have,
\f
Adjoint_{Sp(32)} = 3R \oplus 3V \oplus 3V^{4} \oplus V^{2} \oplus V^{3}
\ff
where $V=R^{9}$ is the vector representation of $Spin(9)$ and 
$V^{p}$ is the antisymmetric $p$-fold product.  The three vectors
are then represented by $\Gamma^{i}, \Gamma^{\# i}$ and $\Gamma^{0i}$,
and the scalars by $\Gamma^{0}, \Gamma^{\#}$ and $\Gamma^{0 \#}$.
These live, respectively, in the vector and trace parts of $X^{AB}$ and 
$B^{AB}_{\pm}= B^{AB}\pm C^{AB}$.  

To see what spin content to expect from the theory we may 
consider the decomposition of
the symmetric $16\times 16$ tensor:
\f
X^{AB}= \delta^{AB}R + \Gamma^{AB}_{i}V^{i} + \Gamma^{AB}_{i_{4}}V^{i^{4}}
\label{symm}
\ff
where we use a notation $i_{p}= [i_{1}\ldots i_{p}]$ for the 
antisymmetric combination.  The antisymmetric tensor is
\f
Y^{AB}= \Gamma^{AB}_{ij}V^{ij} + \Gamma^{AB}_{i_{3}}V^{i^{3}}
\ff
The three scalar's, three vectors and the $V^{ij}$ parameterize
the embedding of the $11$ dimensional DeSitter algebra,
$SO(10,2)$ in $Sp(32)$.  The remaining $V^{3}$ and the three
$V^{4}$'s represent elements of $Sp(32)$ that do not come from
$SO(10,2)$.  In the contraction of $Osp(1|32)$ that becomes
the $11$ dimensional Super-Poincare algebra they become the
central charges.

The $32$ supersymmetry charges decompose  into two
$16$'s of $Spin(9)$ which may be parameterized as
\f
\epsilon_{A}\hat{Q}^{A} + \chi_{A^{\prime}}\hat{Q}^{A^{\prime}}= 
 \left (
\begin{array}{ccc}
    0 & 0 & \epsilon  \\
    0 & 0 & \chi  \\
   \chi^{T} & -\epsilon^{T} & 0
\end{array}
\right ) 
\ff

We now introduce a time coordinate in the direction parameterized
by $\Gamma^{0}$ by the usual trick\cite{} of a matrix compactification.
This means that we expand around a background given by
all fields vanishing except
\f
B^{AB}_{ab} = -C^{AB}_{ab}= \delta^{AB}{\cal D}_{0ab}
\label{tbackground}
\ff
where ${\cal D}_{0}$ has the property as an $N\times N $ matrix that
as $N$ tends to infinity
\f
Tr W [{\cal D}_{0} , M ] = 
{1\over T} \oint dt Tr \left \{ W(t)( \imath {\partial M(t) \over \partial t} +
[ A_{0} , M])  \right \} 
\label{derivative}
\ff
This is done by breaking each $N\times N$ matrix, $M_{AB}$ up into
a very large number $2F+1$ of $P\times P$ blocks, 
$\tilde{M}_{\tilde{a}\tilde{b}}(n)$ such that
\f
Tr_{N\times N } M = \sum_{n=-F}^{F} Tr_{P\times P }\tilde{M}(n)
\ff
${\cal D}_{0ab}$ is then defined as ${\cal D}_{0ab}=K_{ab}+A_{0ab}$
where
\f
Tr W [K,M] = \sum_{n} Tr_{P\times P }W(n) n \tilde{M}(n)
\label{exact}
\ff
If we introduce a fundamental scale $l_{Pl}$ then we can
fourier transform to define
\f
M(t) =  \sum_{n=-F}^{F}e^{i2\pi n t /T} \tilde{M}(n)
\ff
where $T=(2F+1)l_{Pl}$, leading in the limit $F\rightarrow \infty$,
with $T$ held fixed and $l_{Pl}\rightarrow 0$ to (\ref{derivative}).

It is easy to see that (\ref{tbackground}) is a solution to the
equations of motion (\ref{eom}).   Note that all dimensional
quantities will be proportional to some power of $l_{Pl}$,
by definition. The notion of a dimensional scale is just a convenient
device to compare with known physics, the physics actually depends only
on $P$ which we interpret as the ratio of the compactification scale
and $l_{Pl}$.  Since in physics  we expect both the compactification 
scale and 
$l_{Pl}$ to be finite we  regard expressions such as (\ref{derivative})
as shorthand for the more exact expressions of the form of
(\ref{exact}) with $F$ very large but finite.

After some algebra the cubic action (\ref{cubicaction}) takes the form
\begin{eqnarray}
I &=& {1\over T} \oint dt Tr \left \{ \Phi_{A}[\partial_{0}, 
\Phi^{A} ] + \Psi_{A}[\partial_{0}, \Psi^{A} ] 
+  X_{AB}[\partial_{0}, B^{AB}_{+} ]  \right. \nonumber \\
&& \left. -{\cal H} [X,B_{+},\Psi, \Phi] 
+ A_{0ab}{\cal G}^{ab} + {\cal C}(Y,B^{-};X,B_{+}, \Psi,\Phi)
\right \}  
\label{hamform}
\end{eqnarray}
We have made several redefinitions,
\f
C^{AB}= -\delta^{AB}{\cal D}_{0}+ \tilde{C}^{AB}, \ \ 
B^{AB}= + \delta^{AB}{\cal D}_{0}+ \tilde{B}^{AB}. 
\ff
and 
\f
B^{AB}_{\pm}=  \tilde{B}^{AB}\pm \tilde{C}^{AB}.
\ff

We see that the fields have split into a dynamical set,
consisting of $X,B_{+},\Psi,\Phi$ and a remaining non-dynamical
set consisting of $Y$ and $B^{-}$.   $X^{AB}$ are the momenta
conjugate to $B^{AB}_{+}$, while the $B^{AB}_{-}$ are constrained
fields.  
The hamiltonian density in terms of the dynamical fields is,
\f
{\cal H} [X,B_{+},\Psi, \Phi ] =
- B_{AB}^{+}
\left (  \{\Phi^{A},\Phi^{B}\} - \{\Psi^{A},\Psi^{B}\} \right ) 
-2X_{AB} \{\Psi^{A},\Phi^{B}\}
- {3\over 2} X[B^{+},B^{+}]
\ff
The Gauss's law constraint is
\f
{\cal G}^{ab}= [ \Psi_{A} , \Psi^{A} ] + [ \Phi_{A} , \Phi^{A} ]+ 
[ X_{AB}, B^{AB}_{+} ] 
\ff
This is first class and generates local $GL(N,R)$ transformations 
on the dynamical fields.

We see the very interesting fact that $Y^{AB}$ and
$B_{-}^{AB}$ have no conjugate momenta and are then constrained
in terms of the dynamical fields $X,P,\Psi^{\pm}$.  The potential
energy density
for these constrained fields is
\begin{eqnarray}
{\cal C}(Y,B^{-};X,B_{+},\Psi, \Phi )&=&   \{
6Y[X,X] + 2Y[Y,Y] + {3\over 2}X([B_{-},B_{-}]-2[B_{+},B_{-}]
\nonumber \\
 &&+ {1 \over 2} B^{-}_{AB} (  \{\Phi^{A},\Phi^{B}\} - 
 \{\Psi^{A},\Psi^{B}\}  )  \}
\end{eqnarray}

Varying with respect to $Y$ and $B^{-}$, respectively, we have constraint
equations
\f
E^{[AB]}= [X^{[A}_{C},X^{B]C}] +  [Y^{[A}_{\ C},Y^{B]C}]  =0
\label{EAB}
\ff
\f
J^{(AB)}= {1\over 2} \left (  \{\Phi^{A},\Phi^{B}\} 
+ \{\Psi^{A},\Psi^{B}\} \right ) -3[X,B_{-}]+3[X,B_{+}] =0
\label{JAB}
\ff
These define quadratic surfaces in the space of matrices, and 
can be solved to express $Y^{AB}$ and $B^{AB}_{-}$
in terms of the dynamical fields $X,B_{+},\Psi, \Phi $.  The result
is that the total hamiltonian density is
\f
{\cal H}^{total} (X,P,\Psi^{\pm}) = {\cal H} [X,B_{+},\Psi, \Phi]
- {\cal C}(Y(X),B^{-}(X,B_{+},\Psi, \Phi),X,B_{+},\Psi, \Phi)
\ff
This has not yet been done explicitly. Unless there is some miracle
the result will be 
non-polynomial, but this is not surprising given that this is the
case also for general relativity for most choices of variables.
The further analysis of the hamiltonian theory requires careful consideration
of the space of solutions of the constraints, which has not yet
been carried out.

\section{Compactification to an $SO(9,1)$ covariant theory}

We next study a compactification which breaks the symmetry
down to the $D=9+1$ superPoincare algebra.  To do this we
compactify in the $11$'th dimension, which is the degree of
freedom generated by $\Gamma^{\#}$.  We do this by writing
\f
B^{AB}= \delta^{AB}({\cal D}_{\#}-T+b_{+})+\tilde{B}^{AB} ; \ \ 
C^{AB}= \delta^{AB}({\cal D}_{\#}+T+b_{+})+\tilde{C}^{AB}
\label{10+1}
\ff
where $\tilde{B}^{AB}$ and $\tilde{C}^{AB}$ are now tracefree.
We expand around a classical solution in which all fields except
${\cal D}_{\#}$ vanish and we 
impose conditions on ${\cal D}_{\#}$ identical to those
imposed in the last section on ${\cal D}_{0}$.  
$b_{+}$
carries the fluctuations around the compactification radius.
$T$ is the field in the direction $\Gamma^{0}$.  

The cubic action is now most simply expressed in terms of redefined fields,
$\tilde{B}_{\pm}^{AB}=\tilde{B}^{AB} \pm \tilde{C}^{AB}$,
$X^{AB}=\tilde{X}^{AB} + \delta^{AB}x$ and 
$\Phi_{\pm}^{A}= \Phi^{A}\pm \Psi^{A}$.  We have
\f
I^{2}_{\#}= -Tr\left \{
\Phi_{A}^{+}[{\cal D}_{\#}, \Phi^{A}_{-}]+ 
+12\tilde{B}_{AB}^{-}[{\cal D}_{\#}, \tilde{X}^{AB}]
 +12x[{\cal D}_{\#}, T] - {\cal H}_{\#}+{\cal C}_{\#}
\right \}
\ff
where the hamiltonian is now
\begin{eqnarray}
{\cal H}_{\#}&=&{1\over 4}\tilde{B}^{-}
\left (\{\Phi^{+},\Phi^{+}\}+\{\Phi^{-},\Phi^{-}\} \right )
+{1\over 2}\tilde{X} \left (\{\Phi^{+},\Phi^{+}\}-\{\Phi^{-},\Phi^{-}\}
-2\{\Phi^{+},\Phi^{-}\}\right )
\nonumber \\ 
&&-{3\over 2} X[B_{-},B_{-}]
-{1\over 2} T \left (\{\Phi^{+},\Phi^{+}\}+\{\Phi^{-},\Phi^{-}\} \right )
+{1\over 2}x 
\left ( \{\Phi^{+},\Phi^{+}\}-\{\Phi^{-},\Phi^{-}\}-2\{\Phi^{+},\Phi^{-}\} \right )
\end{eqnarray}
and the constraints come from
\f
{\cal C}_{\#}= {2\over 3}X\left ([B^{+},B^{+}] +2 [B^{+},B^{-}] \right )
    + {1\over 2}B^{+}\{\Phi^{+},\Phi^{-}\}  +2Y[Y,Y]+6Y[X,X]
\ff

The quadratic term tells us how to perform the quantization with
respect to the Euclidean time $X^{\#}$.  
We see that we again have a division into dynamical and non-dynamical
fields.  
The dynamical fields now are $\tilde{B}_{-},\tilde{X},x,T,\Phi^{\pm}$.
The non-dynamical fields are $B_{+},b_{+}$ and
$Y^{AB}$.  These will be determined by constraints analogous to
(\ref{EAB}) and (\ref{JAB}) as a result of which we will have
\f
Y^{AB}=Y^{AB}[X] ; \ \ \ 
B^{AB}_{+}= B^{AB}_{+}[\tilde{B}_{-},\tilde{X},x,T,\Phi^{\pm}]
\ff
We note that these constrained fields 
make up an $SO(9,1)$ scalar,$b_{+}$, 
two form, $V^{\mu\nu}$, and
four form, $W^{\mu \nu \lambda \sigma}$ 
(with $9+1$ dimensional indices $\mu,\nu=(0,i)$). 
These are given, in terms of $9D$ gamma matrices, by
\f
B_{+}^{AB}=\gamma^{AB}_{i}V^{0i}+ \gamma^{AB}_{i_{4}}W^{i_{4}}
\ff
\f
Y^{AB}=\gamma^{AB}_{ij}V{ij}+ \gamma^{AB}_{i_{3}}W^{0i_{3}}
\ff 

We will not here solve the constraints and compute the resulting
hamiltonian for the unconstrained fields. As a result, we cannot
commute the one-loop effective potential of the dynamical
fields precisely.  But we can use the unbroken symmetry to 
determine its form.
We first organize the dynamical fields in terms
of $9+1$ dimensional tensors.  The $T$ component combines with 
$X^{i}$, where 
\f
X^{AB}=\gamma^{AB}_{i}X^{i}+ \gamma^{AB}_{i_{4}}X^{i_{4}}
\ff
to make the $9+1$ vector of matrices.
\f
X^{\mu}=(T,X^{i})
\ff
The remaining fields are the $X^{i_{4}}$, the fields in
$B^{AB}_{-}$, given by 
\f
B_{-}^{AB}=\gamma^{AB}_{i}B_{-}^{i}+ \gamma^{AB}_{i_{4}}B_{-}^{i_{4}}
\ff
and the scalar $x$.  These do not combine to form any more 
$SO(9.1)$ tensors, although they play the role of canonical momenta
(in the ${\cal D}_{\#}$ time) to fields that are parts of
$SO(9,1)$ tensors. They must then be eliminated in the computation
of the one loop effective potential, which then will have,
at least to lowest order in the fields, the 
$SO(9,1)$ invariant form\cite{IKKT}
\f
S^{\#}= Tr \left \{ [X^{\mu}, X^{\nu}][X{\mu}, X_{\nu}] + 
\Psi_{\bar{\alpha}}[X^{\mu}, \Psi_{\bar{\beta}} ] 
\Gamma_{\mu}^{\bar{\alpha}\bar{\beta}} \right \} 
\ff
Here $\bar{\alpha},\bar{\beta}=(A,A^{\prime})$ is a $32$ component
$SO(9,1)$ spinor index and 
$\Psi_{\bar{\alpha}}=(\Psi^{A},\Phi^{A^{\prime}})$.
The spinor may be decomposed into chiral eigenstates
$\Psi_{\bar{\alpha}}^{\pm}=(\Psi^{A},\pm \Psi^{A^{\prime}})$.
Under supersymmetry transformations generated by
$Q^{\pm}_{A}=Q^{1}_{A}\pm Q^{2}_{A}$ we have 
\f
\delta \Psi^{\pm}_{A}= \pm {\cal D}_{\#}Q^{\pm}_{A}
\ff
This tells us that if we keep both fermion fields in the
dynamics, we have two rigid supersymmetries, with
$[{\cal D}_{\#}, Q^{\pm}_{A}]=0$.  However, if we 
decouple one of the fields, say $\Psi^{+}$ then we
need only require $[{\cal D}_{\#}, Q^{-}_{A}]=0$ so the theory
will be invariant under one global and one local supersymmetry.
This suggests that supersymmetry will protect one, but not both
spinor fields, so we are left with the field content of the
IKKT model\cite{IKKT}.  
It is not hard to see that if we ignore the constrained fields
completely the one-loop effective potential is exactly of this form.
But a precise calculation cannot be done until the constraints
have been properly dealt with.

\section{Triple compactification and the discrete light cone
quantization}

We next consider a different compactification, which is suitable
for extracting the infinite momentum frame description of the
theory in $10+1$ dimensions.  This should have as the explicit
symmetry only the super-Euclidean group in $9$ dimensions.  
To construct this limit we study a triple compactification
on all three of the $Spin(9)$
scalar modes of $Z_{\alpha}^{\beta}$.  These correspond to
$X^{\pm}= X^{0}\pm X^{\#}$ and the longitudinal boost
$\Gamma^{0\#}$.  The idea is then to compute the one loop
effective potential that follows from integrating out the
modes of the fields in the time coordinate generated by $\Gamma^{0\#}$.
This gives a theory expressed in terms of $SO(9)$ transverse
degrees of freedom and the light cone coordinates and
momenta defined in terms of $X^{\pm}$.  

Again we cannot make an exact calculation as there are
constraints in the $\Gamma^{0\#}$ time, analogous to those
we encountered before.  But we can use group theory to
constrain the possible form of the one-loop effective potential,
and we can also 
verify that the terms in it do appear in a version of the calculation in
which the constrained degrees of freedom are ignored rather than
solved for.

We begin by compactifying only the longitudinal boost direction,
which is given by the background in which all fields vanish except
\f
X^{AB}=\delta^{AB}{\cal D}_{\tau}
\ff
where ${\cal D}_{\tau}$ is defined similarly to 
${\cal D}_{0}$ above.  We find an expression similar to the
previous one, differing of course because we are introducing
a different time coordinate,
\begin{eqnarray}
I &=& - {1\over \tilde{T}} \oint d\tau Tr \left \{ 
\Psi_{A}[{\cal D}_{0}, \Phi^{A} ] +  B_{AB}[{\cal D}_{0}, C^{AB} ]  \right. 
\nonumber \\
&& \left. -{\cal H}_{\tau} [B,C,\Psi, \Phi] + 
+ A_{\tau ab}{\cal G}^{ab}_{\tau} + {\cal C}^{\tau}(Y,X,\mbox{otherfields})
\right \}  
\label{hamtau}
\end{eqnarray}
We find the unconstrained hamiltonian density is now simply
\f
{\cal H}_{\tau} [B,C,\Psi, \Phi]= B_{AB}\{\Phi^{A},\Phi^{B}\} -
C_{AB} \{\Psi^{A},\Psi^{B}\}
\ff
Now it is $X^{AB}$ along with $Y^{AB}$ which is to be determined
by the solution to constraints.  The new constrained potential energy is 
\f
{\cal C}^{\tau}(Y,X,\mbox{otherfields})= 6\tilde{X}[B,C]
-2 Y[Y,Y] -6 X[Y,Y] +2\tilde{X}_{AB} \{\Psi^{A},\Phi^{B}\}
\ff
We first consider what happens if we simply ignore these
constrained fields, $X^{AB}$ and $Y^{AB}$ and study the
theory defined by (\ref{hamtau}) with the term
${\cal C}^{\tau}$ ignored.  The dynamical fields are
only $C^{AB}, B^{AB},\Phi^{A},\Psi^{A}$.  
Keeping in mind the fact that $\tau$ is a Euclidean
time coordinate, we can integrate out over the
modes which propagate in $\tau$.   The effective potential
for the unconstrained fields 
is then, to lowest order, of the form, 
\f
I=I^{0}+\hbar I^{1}
\ff
where
\f
I^{0}=Tr {\cal H}_{\tau}
\ff
and the one loop effective potential has the form
\f
I^{1}= Tr \left \{
[B,C]^{2}+\Phi[B,\Phi] + \Psi [C, \Psi ]
\right \}
\ff

We next compactify the $x^{+}$ and $x^{-}$ directions, which
are generated by $\Gamma^{\pm} =\Gamma^{0}\pm \Gamma^{\#}$.
We do this by writing
\f
B^{AB}= \delta^{AB}{\cal D}_{-}+ \tilde{B}^{AB} , \ \ 
C^{AB}= \delta^{AB}{\cal D}_{+}+ \tilde{C}^{AB}
\ff
and expand around the background whose only non-zero fields
are ${\cal D}_{\tau}, {\cal D}_{+}, {\cal D}_{-}$, where
${\cal D}_{+}, {\cal D}_{-}$ are defined as 
in the cases of the other time coordinates.  The compactification 
radii are $R^{\pm}$.  The result is a
$1+1$ field theory defined on the torus, whose effective action 
contains the terms
\begin{eqnarray}
S&=&{1 \over R^{+}R^{-}} \oint \oint dx^{-}dx^{+}   
\left \{ [{\cal D}_{-},\tilde{C}]^{2}+ [{\cal D}_{+},\tilde{B}]^{2}
+\Phi [{\cal D}_{-}, \Phi ] +\Phi [{\cal D}_{+}, \Phi ]
+\Psi [{\cal D}_{-}, \Psi ] +\Psi [{\cal D}_{+}, \Psi ]
\right.  \nonumber \\
&& \left. +[\tilde{B},\tilde{C}]^{2} +
\Phi [\tilde{B}, \Phi ] +\Phi [\tilde{C}, \Phi ]
+\Psi [\tilde{B}, \Psi ] +\Psi [\tilde{C}, \Psi ]
\right \}
\end{eqnarray} 
We next perform a very large boost in the positive $\#$ direction,
which in the limit will take us to the infinite momentum frame. 
In the limit all terms proportional to ${\cal D}_{-}$ decouple
as those backwards moving modes  have in the limit infinite energy.
The degrees of freedom which
survive the limit are only those with kinetic energies
proportional to ${\cal D}_{+}$, their dynamics 
is described by the action containing the terms,
\begin{eqnarray}
I^{+}&=& {1\over R^{+}} \oint dx^{+}  Tr \left \{  
\Phi_{A} [{\cal D}_{+}, \Phi^{A} ]  +
\Psi_{A} [{\cal D}_{+}, \Psi^{A} ]
 +[{\cal D}_{+}, \tilde{B}_{AB}][{\cal D}_{+}, \tilde{B}^{AB}] \right.
 \nonumber \\
 &&\left. + \Phi_{A} [\tilde{B}^{AB}, \Phi_{A} ] 
 +  \Psi_{A} [\tilde{B}^{AB}, \Psi_{A} ]
\right \}
\label{almostIMF}
\end{eqnarray}
This is close to the standard $DLCQ$ action for $\cal M$ theory.
There is again one fermion field too many for there to remain
a local supersymmetry. When we integrate this out 
this leaves us with an effective action of the form
\f
I^{IMF}= {1\over R^{+}} \oint dx^{+}  Tr \left \{    
\Phi_{A} [{\cal D}_{+}, \Phi^{A} ]
 +[{\cal D}_{+}, \tilde{B}_{AB}][{\cal D}_{+}, \tilde{B}^{AB}]
 + \Phi_{A} [\tilde{B}^{AB}, \Phi_{A} ] 
 +[\tilde{B}^{AB},\tilde{B}^{CD}][\tilde{B}_{AB},\tilde{B}_{CD}]
\right \}
\label{IMF}
\ff

Our infinite momentum frame action (\ref{IMF}) is almost, but not quite
the matrix model for $\cal M$ theory described in \cite{CH,dWHN,BFSS}
which is simultaneously a description of the supermembrane
in light cone gauge and the reduction to one dimension of
$D=10$ supersymmetric Yang-Mills theory.  The difference
is that the tracefree part of the $B^{AB}$ field contains
a five form as well as a vector, which is given by the
decomposition (\ref{symm}) of the symmetric trace free
spinor $B^{AB}$.  Thus the theory is an extension of the
usual matrix model with 
\f
X^{i}\gamma^{AB}_{i} \rightarrow B^{AB}= 
\gamma^{AB}_{i}X^{i} + \gamma^{AB}_{ijkl}V^{ijkl}
\ff
The additional degree of freedom may be interpreted
to be a transverse
five-form field $A_{jklmn}=V^{i_{4}*}_{jklmn}$.
The chief consequence of its addition is that
the supersymmetry algebra now contains 
central terms.    This
will be discussed in more
detail elsewhere.

\section{Conclusions}

What we have reported here is just the first step in the 
analysis of the model given by the cubic action.  The most important
technical problem to be resolved is the correct way to handle
the constraints which arise in the different quantizations.
Once this is done the effective action can be calculated exactly
to any order desired, and the results compared with the IKKT and
dWHN-BFSS forms of the matrix theory.  What we have argued here
is that by expanding around the appropriate classical solutions
those theories will be reproduced, with the possible addition
of a transverse five form field in the light cone gauge case.

If the theory passes this test then it will be of interest to
investigate whether all the known consistent perturbative string
theories, together with the web of dualities, may be understood
as arising from expanding the present model around different
classical solutions.  It is known that the several different
string theories can be gotten by compactifying the IKKT
and dWHN-BFSS matrix models\cite{BFSS,moreBFSS,IKKT};
it will be of interest to see if others may be found. It would also
be interesting to see if there are compactifications of this theory
which reduce to the proposals presented in 
\cite{andy,park}\footnote{I would like to thank Miao Li for pointing 
out to me the latter work.}.

Another set of questions to explore arise from the relationship
between the simplest symplectic matrix model and Chern-Simons theory
we described in section 2.  This suggests that the triple 
compactification of the $Osp(1|32)$ theory, 
one limit of which we argued gives rise to the light cone gauge
matrix model, may be studied also as a $2+1$ dimensional
topological quantum field theory.  
A closely related set of structures are the basis of the 
connection between this model and the background 
independent approaches to membrane and $\cal M$ theory described
in \cite{mpaper,tubes,pqtubes}. This will be discussed elsewhere.

Beyond this there are several deep questions. The first is
the question of what the right quantization procedure should
be for the full theory. In this model time is only introduced
by expanding around a classical solution, given by an
appropriate compactification.  It is not at all clear if a quantum
theory can be defined in the absence of any time variable, for in that
case there is no canonical formulation to base the quantization on.

It is possible that there may be an unconvential answer to this question,
in which quantum statistics emerges for the local observables when the
matrices are thermalized.  
This is suggested by the fact that in the absence of the choice of
a time variable no clear distinction can be made between thermal
and quantum fluctuations, as that depends on the signature of the
action.  General arguments tell us
that in quantum gravity the distinction between quantum and thermal
statistics should exist only relative to local inertial reference frames
in spacetimes with lorentzian signature\cite{mixing}.  A matrix
model in which quantum statistics emerged from a large $N$ limit
of ordinary statistics was described in \cite{hidden}.
It was found that such models are
able to evade the experimental limits on local hidden variables 
theories because only the eigenvalues of
the matrices are associated with local observables, 
while the matrix elements themselves are non-local.  
Alternatively it may be that there is an algebraic approach to
the quantization of such systems defined by an appropriate triple
product.  Such theories have been studied by \cite{triple}. 

Another set of
questions arises from the fact that the time coordinates introduced
via compactifications are periodic.  It is of interest to 
understand if this is fundamental or if there are ways to 
introduce time and space coordinates which are not compact. 

Finally, we note that there are a number of other models which might
be studied, which have many features in common with the present one.
By complexifying the degrees of freedom of our model we may arrive
at a model based on $SU(16,16|1)$.  It is interesting to consider
this as an extension of twistor theory, as that
is based on an $SU(2,2)$ symmetry.

\section*{Acknowledgments}

I would like to thank Chris Isham, Jerome Gauntlett, 
Chris Hull, Miao Li, Yi Ling, Fotini Markopoulou, Djordje Minic,
Michael Reisenberger, Steve Shenker, Kelly Stelle and Paul Townsend for helpful
discussions and correspondence during the course of this work.
I am very grateful to Chris Isham for the hospitality of the Theoretical
Physics Group at Imperial College, where this work was carried out. 
This work was also supported by the NSF through grant
PHY95-14240, by PPARC through SPG grant 613, and 
a gift from the Jesse Phillips Foundation.

\end{document}